\documentclass[12pt]{article}
\def\hybrid{\topmargin 0pt      \oddsidemargin 0pt
        \headheight 0pt \headsep 0pt
       \textwidth 6.5in        
        \textheight 9in         
        \marginparwidth .875in
        \parskip 5pt plus 1pt   \jot = 1.5ex}
\catcode`\@=11
\def\marginnote#1{}
\newcount\hour
\newcount\minute
\newtoks\amorpm
\hour=\time\divide\hour by60
\minute=\time{\multiply\hour by60 \global\advance\minute
by-\hour}\edef\standardtime{{\ifnum\hour<12
\global\amorpm={am}%
        \else\global\amorpm={pm}\advance\hour by-12 \fi
        \ifnum\hour=0 \hour=12 \fi
        \number\hour:\ifnum\minute<10
0\fi\number\minute\the\amorpm}}
\edef\militarytime{\number\hour:\ifnum\minute<10
0\fi\number\minute}

\def\draftlabel#1{{\@bsphack\if@filesw {\let\thepage\relax
   \xdef\@gtempa{\write\@auxout{\string
      \newlabel{#1}{{\@currentlabel}{\thepage}}}}}\@gtempa
   \if@nobreak \ifvmode\nobreak\fi\fi\fi\@esphack}
        \gdef\@eqnlabel{#1}}
\def\@eqnlabel{}
\def\@vacuum{}
\def\draftmarginnote#1{\marginpar{\raggedright\scriptsize\tt#1}}
\def\draft{\oddsidemargin -.5truein
        \def\@oddfoot{\sl preliminary draft \hfil
        \rm\thepage\hfil\sl\today\quad\militarytime}
        \let\@evenfoot\@oddfoot \overfullrule 3pt
        \let\label=\draftlabel
        \let\marginnote=\draftmarginnote

\def\@eqnnum{(\theequation)\rlap{\kern\marginparsep\tt\@eqnlabel}%
\global\let\@eqnlabel\@vacuum}  }

\def\numberbysection{\@addtoreset{equation}{section}
        \def\theequation{\thesection.\arabic{equation}}}

\def\underline#1{\relax\ifmmode\@@underline#1\else
 $\@@underline{\hbox{#1}}$\relax\fi}

\catcode`@=12
\relax
\numberbysection
\hybrid
\def\r2{\sqrt{2}}
\def\beq{\begin{equation}}
\def\eeq{\end{equation}}
\def\bea{\begin{eqnarray}}
\def\eea{\end{eqnarray}}

\def\fin{\end{document}}
\begin{document}
\def\nom{}
\def\modification{} 
\begin{titlepage}
\nopagebreak
\nom\modification
\begin{flushright}
LPTENS-99/13\\
hep--th/9903218
\\
March   1999
\end{flushright}
\vglue 2.5 true cm
\begin{center}
{\large
LAX EQUATIONS \\
\medskip 
  IN TEN DIMENSIONAL SUPERSYMMETRIC\\ 
\medskip\smallskip  
CLASSICAL YANG--MILLS THEORIES}\\
\vglue 1 true cm
{\itshape \itshape Contribution\footnote{To appear in a volume published by the Max Planck Institute
for Mathematics Bonn.} to the International Seminar on Integrable systems\\ In Memoriam Mikhail V. 
Saveliev} \\
\vglue 1 true cm
{ Jean--Loup GERVAIS}\\ {\footnotesize Laboratoire de Physique Th\'eorique de
l'\'Ecole Normale Sup\'erieure
\footnote{UMR 8548:  Unit\'e Mixte du Centre National de la Recherche Scientifique, et de 
l'\'Ecole Normale Sup\'erieure. },\\ 24 rue Lhomond, 75231 Paris C\'EDEX
05, ~France.}
\end{center}
\bigskip 
\baselineskip .4 true cm
\noindent
\begin{abstract}
In a recent paper (hep-th/9811108), Saveliev and the author showed that there exits an on-shell
light cone gauge where   the  non-linear part  of the field equations reduces to a (super) version
of Yang's equations which may be solved by methods inspired by the ones previously developed for
self-dual Yang-Mills equations in four dimensions.  Here, the analogy between these latter
theories  and the present ones is pushed further by writing down a set of super partial linear
differential equations whose consistency conditions may be derived from the  SUSY Y-M equations in 
ten dimensions, and which are the analogues of the Lax pair of Belavin and Zakharov. On the simplest
example of the two pole ansatz, it is shown that the same solution-generating techniques are at
work, as for the derivation of the celebrated multi-instanton solutions carried out in the late 
seventies. The present Lax representation, however, is only a consequence of 
(instead of being equivalent to) the field equations, in contrast with  the  Belavin Zakharov Lax
pair.

\end{abstract}

\vfill
\end{titlepage}
\section{Introduction} 
{  \itshape 
 The general solution of spherically symmetric self-dual Yang-Mills equations discovered
by Lesnov and Saveliev two decades ago has led to extraordinary developments. I met Misha for the
first time in 1992 when this work had already proven to be so important for two dimensional
conformal/integrable systems. We immediately started to collaborate and have done
so  ever since. Unlike
many of his country men he felt that he should not leave his country for good, and  fought for
his family and him-self while keeping a remarkable enthousiasm for research.  Working with Misha 
has been a wonderful experience which terminated so abruptly! I will always remember our excited
and friendly discussions, his kindness and enthousiam, his fantastic knowledge of the scientific
literature!   We, at
\'Ecole Normale,  were lucky enough  to invite him for several extended visits which were extremely
fruitful. Misha and I also met in other places, but altogether much too rarely, and exchanged
uncountably many email messages. Now I am sorry for the many occasions to meet him that I had to
decline. In particular I never found time to visit him in Russia.   Our  collaboration on the
present  subject was entirely by e-mail. Our last encouter in person was in Cambridge (UK) at the
beginning of March 1997. At that time I thought as a matter of course that we would meet soon again,
but this is not so! In the large number of email messages we exchanged since then, it is clear that
he was under a great pressure, but yet he was always coming up with exciting ideas, calculations and
so on.

My other regret is that, although we were very good friends, we seldom had time to socialise
outside  research. I will always remember these few very warm and friendly encounters, and
especially when his  Svetlana's (as he used to say)  were present.   

 M. Saveliev  was great both as a scientist and as a
human being. He was obviously such a good father,  husband, friend!}

\vglue 1 true cm 
In recent  times we turned\cite{GS98} to the classical integration of theories in more than two
dimensions with local extended supersymmetries. Our motivation was twofold. On the one hand this
problem is very important for the recent developments in  duality and M theory. On the other hand, 
the recent advances initiated by Seiberg and Witten indicate that  these theories  are in many ways 
higher dimensional analogues of two dimensional conformal/integrable systems, so that progress may be
expected. Since fall 1997, we have studied  super Yang-Mills theories in ten dimensions. There, it
was shown by Witten\cite{W86} that the field equations are equivalent to  flatness conditions. This
is   a priori  similar to well known basic ones  of Toda theories, albeit no real progress could
be made at that time, since the corresponding Lax type equations  involve an arbitrary light like
vector which plays the role of a spectral parameter.  At first, we reformulated the field
equations  in a way which is similar to    a super version  of    the higher dimensional
generalisations of Toda theories developed by  Razumov and Saveliev\cite{RS97},  where the
Yang-Mills gauge algebra is extended to a super one. This has not yet been published since,
contrary to our initial hope, the two types of theories do not seem to be equivalent. I hope to
return to this problem in a near future.  In the mean time, we found the existence of an on-shell
gauge, in super Yang-Mills where the field equations simplify tremendously and where the first
similarity with self-dual Yang-Mills in four dimensions came out\cite{GS98}. This directly led to
the present progress.

As is well known, super Yang-Mills theories in ten dimensions just describes a standard non abelian
gauge field coupled with a charged Majorana-Weyl spinor field in the adjoint representation of
the gauge group. The dynamics
is thus specified by the standard action 
\beq
S=\int d^{10} x {\> \rm Tr   }
\left\{
-{1\over 4}Y_{mn}Y^{mn}
+{1\over 2}\bar \phi\left(\Gamma^m \partial_m \phi+\left[X_m,\, \phi\right]_- \right)\right\}, 
\label{action}
\eeq
\beq
Y_{mn}=\partial_mX_n-\partial_nX_m +\left[X_m,\, X_n\right]_-.
\label{F0def} 
\eeq
The notations are as follows\footnote{They are  essentially the same as in ref.\cite{GS98}.}: 
$X_m(\underline x)$ is the vector potential, $\phi(\underline x) $ is the Majorana-Weyl spinor. Both
are matrices in the adjoint representation of the gauge group 
${\bf G}$.  Latin indices
$m=0,\ldots 9$ describe Minkowski components.  Greek indices $\alpha=1,\ldots 16$ denote
chiral spinor components. We will use the superspace formulation with odd coordinates
$\theta^\alpha$. The  super vector potentials, which are valued in the gauge group, are noted  
$A_m\left(\underline x,\underline \theta\right)$, $A_\alpha\left(\underline x,\underline
\theta\right)$. As shown in refs. \cite{W86}, \cite{AFJ88}, we may
remove all the additional fields and uniquely reconstruct the physical fields $X_m$, $\phi$ from
$A_m$ and $A_\alpha$ if we impose the condition $\theta^\alpha A_\alpha=0$ on the latter.

With this condition, it was shown in refs. \cite{W86}, \cite{AFJ88}, that the field equations
derived from the Lagrangian  \ref{action} are equivalent to the flatness conditions 
\beq
{\cal F}_{\alpha \beta=0}, 
\label{flat}
\eeq
where ${\cal F}$ is the supercovariant curvature 
\beq
{\cal F}_{\alpha \beta}=D_\alpha A_\beta+D_\beta A_\alpha+\left[A_\alpha,\, A_\beta\right]+
2\left(\sigma^m\right)_{\alpha\beta}A_m.  
\label{curdef}
\eeq
 $D_\alpha$ denote the superderivatives
\beq
D_\alpha=\partial_\alpha-\left(\sigma^m\right)_{\alpha \beta} 
\theta^\beta {\partial_m}, 
\label{sddef}
\eeq
and we use the Dirac matrices 
\beq
\Gamma^m=\left(\begin{array}{cc}
0_{16\times16}&\left(\left(\sigma^m\right)^{\alpha\beta}\right)\\
\left(\left(\sigma^m\right)_{\alpha\beta}\right)&0_{16\times16}
\end{array}\right),\quad  
\Gamma^{11}= \left(\begin{array}{cc}
1_{16\times16}&0\\0&-1_{16\times16}\end{array}\right).
\label{real1}
\eeq
Throughout the paper, it will be convenient to use the following particular realisation: 
\beq
\left(\left(\sigma^{9}\right)^{\alpha\beta}\right)=
\left(\left(\sigma^{9}\right)_{\alpha\beta}\right)=
\left(\begin{array}{cc}
-1_{8\times 8}&0_{8\times 8}\\
0_{8\times 8}&1_{8\times 8}
\end{array}\right)
\label{real2}
\eeq
\beq
\left(\left(\sigma^{0}\right)^{\alpha\beta}\right)=-
\left(\left(\sigma^{0}\right)_{\alpha\beta}\right)=
\left(\begin{array}{cc}
1_{8\times 8}&0_{8\times 8}\\
0_{8\times 8}&1_{8\times 8}
\end{array}\right)
\label{real3}
\eeq
\beq
\left(\left(\sigma^{i}\right)^{\alpha\beta}\right)=
\left(\left(\sigma^{i}\right)_{\alpha\beta}\right)=\left(\begin{array}{cc}
0&\gamma^i_{\mu,\overline \nu}\\
\left(\gamma^{i\, T}\right)_{\nu,\overline \mu}&0
\end{array}\right),\quad  i=1,\ldots 8. 
\label{real4}
\eeq
The convention for greek letters is as follows: Letters from the beginning of the alphabet run from
1 to 16. Letters from the middle of alphabet run from 1 to 8. In this way,  we shall separate
the two spinor representations of $O(8)$ by rewriting $\alpha_1,\ldots, \alpha_{16} $  as 
$\mu_1,\ldots, \mu_8, \overline \mu_1,\ldots, \overline \mu_8$
   
Using the above explicit realisations on sees that the equations to solve take the form  
\begin{eqnarray}
F_{\mu \nu}\equiv D_\mu A_\nu+D_\nu A_\mu +\left[A_\mu,\,
A_\nu\right]_+&=&2\delta_{\mu\nu}\left(A_0+A_9\right)\label{dynuu}\\
F_{\overline \mu \overline \nu}\equiv  D_{\overline \mu} A_{\overline \nu}+D_{\overline \nu}
A_{\overline \mu} +\left[A_{\overline \mu},\, A_{\overline \nu}\right]_+&=&
2\delta_{{\overline \mu}{\overline \nu}}\left(A_0-A_9\right)\label{dyndd}\\  
F_{ \mu \overline \nu}\equiv   D_{ \mu} A_{\overline \nu}+D_{\overline \nu} A_{ \mu}
+\left[A_{\mu},\, A_{\overline \nu}\right]_+&=&-2\sum_{i=1}^8 A_i\gamma^i_{\mu,\overline
\nu}\label{dynud}
\end{eqnarray}
In my last  paper with M. Saveliev \cite{GS98},  these flatness conditions in superspace were used 
to go to an on-shell light-cone gauge where half of the superfields vanish. After reduction to
$(1+1)$ dimensions, the non-linear part of the equations was transformed into equations for a scalar
superfield which are (super) analogues of the so called Yang equations which were much studied in
connection with solutions of self-dual Yang-Mills  equations in four dimensions. The main
differences between the two type of relations is that derivatives are now replaced by
superderivatives, that there are sixteen equations instead of four, and that the indices are paired
differently. Nevertheless, it was found that these novel features are precisely such that the
equations may be solved by methods very similar to the ones developed in connection with self-dual
Yang-Mills in four dimensions.  The aim of the present paper is to push this analogy much further,
by deriving the analogues of the Lax pair of Belavin Zakharov\cite{BZ78} which was instrumental for
deriving multi-instanton solutions at the end of  the seventies. 

\section{The Lax representation}
The original  theory is $O(9,1)$ invariant, but the choice of Dirac matrices just summarized is
covariant only under a particular $O(8)$ subgroup. The Lax reprsentation will come out after
picking up a particular $O(7)$ subgroup of the latter. This done simply by remarking  that we may
choose one $\gamma^i$ to be the unit matrix, in which case the others are antisymmetric and obey
the $O(7)$ Dirac algebra.  This is so, for instance in  the following explicit representation of the
$O(8)$ gamma matrices, where $\gamma^8$ is equal to one, which we will use throughout:   
\begin{eqnarray}
\gamma^1= \tau   _1\otimes \tau   _3\tau   _1\otimes {\bf 1} \quad &\quad 
\gamma^5=\tau   _3\otimes \tau   _3\tau   _1\otimes {\bf 1} \nonumber\\
\gamma^2= {\bf 1}\otimes \tau   _1\otimes \tau   _3\tau   _1 \quad &\quad
\gamma^6= {\bf 1}\otimes \tau   _3\otimes \tau   _3\tau   _1   \nonumber\\
\gamma^3=\tau   _3\tau   _1 \otimes {\bf 1}\otimes \tau   _1 \quad &\quad
\gamma^7= \tau   _3\tau   _1\otimes {\bf 1} \otimes \tau   _3  \nonumber\\
\gamma^4= \tau   _3\tau   _1\otimes \tau   _3\tau   _1\otimes \tau   _3\tau   _1 \quad &\quad
\gamma^8={\bf 1}\otimes  {\bf 1}\otimes {\bf 1}. 
\label{gamdef}
\end{eqnarray}
 With this choice, it follows from equations \ref{dynuu}--\ref{dynud} that 
\beq
F_{\mu \nu}=2\delta_{\mu \nu}\left(A_0+A_g\right),\quad 
F_{\overline \mu \overline \nu}=2\delta_{\overline \mu \overline \nu}\left(A_0-A_g\right),\quad 
F_{\mu \overline \nu}+F_{ \nu \overline \mu}=-4\delta_{ \mu \nu}A_8 . 
\label{symdyn}
\eeq
We have symmetrized the mixed (last) equations so that the right-hand sides only involve 
Kronecker delta's in the spinor indices.  By taking $\gamma^8$  to be the unit
matrix, we have introduced  a  mapping between overlined and non overlined indices.
Accordingly, in the previous equation and hereafter, whenever we write an overlined index and non
overlined one with the same letter  (such as $\mu$ and $\overline \mu$) we mean that they are
numerically equal, so that 
$\gamma^8_{\mu \overline \mu}=1$. Next, in parallel with what was done for self-dual Yang-Mills
in four dimensions, it is convenient to go to complex (super) coordinates. 
Thus we introduce, with $i$ the square root of minus one\footnote{For the new symbols, the group
theoretical meaning of the fermionic indices $\mu$ $\overline \mu$ is lost. We adopt this
convention to avoid clusy notations.},  
$$
G_{\mu \nu}=F_{\mu \nu}-F_{\overline \mu \overline \nu}
+iF_{\overline \mu  \nu}+iF_{\mu \overline \nu} 
$$
$$
G_{\overline \mu \overline \nu}=F_{\mu \nu}-F_{\overline \mu \overline \nu}
-iF_{\overline \mu  \nu}-iF_{\mu \overline \nu}, 
$$
\beq
G_{ \mu \overline \nu}=F_{\mu \nu}+F_{\overline \mu \overline \nu}
+iF_{\overline \mu  \nu}-iF_{\mu \overline \nu},  
\label{Gdef}
\eeq
\beq
\Delta_\mu= D_\mu +iD_{\overline \mu},\quad 
\Delta_{\overline \mu}= D_\mu -iD_{\overline \mu}, 
\label{Deltadef}
\eeq 
\beq
B_\mu= A_\mu +iA_{\overline \mu},\quad  
B_{\overline \mu}= A_\mu -iA_{\overline \mu}.
\label{Bdef}
\eeq
A straightforward computation shows that 
$$
\left[\Delta_\mu,\, \Delta_\nu\right]_+
=4\delta_{\mu \nu}\left(\partial_9-i\partial_8 \right),\quad  
\left[\Delta_{\overline \mu},\, \Delta_{\overline \nu}\right]_+=
4\delta_{\mu\nu}\left(\partial_9+i\partial_8 \right),  
$$
\beq
\left[\Delta_{ \mu},\, \Delta_{\overline \nu}\right]_++
\left[\Delta_{ \nu},\, \Delta_{\overline \mu}\right]_+
=8\delta_{\mu
\nu}\partial_0
\label{anti}
\eeq

Consider, now the system of differential equations
\beq
{\cal D}_\mu\Psi \left(\lambda\right)\equiv 
\left(\Delta_{\mu}+\lambda \Delta_{\overline \mu}+B_{\mu}+\lambda B_{\overline
\mu}\right)\Psi(\lambda)=0,  \mu=1,\ldots, 8. 
\label{BZ}
\eeq
Of course, although we do not write it for simplicity of notations, $\Psi(\lambda)$ is a
superfield function of $\underline x$ and $\underline \theta$. The parameter $\lambda$ is an
arbitrary complex number. The consistency condition of these equations is 
\beq
\left[{\cal D}_\mu,\, {\cal D}_\nu\right]_+\Psi(\lambda)=0. 
\label{cons}
\eeq
This gives 
$$
\left\{4\delta_{\mu \nu}\left(\partial_9-i\partial_8 \right) 
+G_{\mu \nu}\right\}\Psi 
+\lambda\left\{8\delta_{\mu \nu}\partial_0    
+G_{\nu\overline \mu}
+ G_{\overline \mu\nu} \right\}\Psi 
$$
$$
+\lambda^2\left\{ 4\delta_{\mu
\nu}\left(\partial_9+i\partial_8 \right) 
+ G_{\overline \nu\overline \mu}
\right\}\Psi=0. 
$$
Thus we correctly get that, for $\mu\not=\nu$
$$
G_{\mu \nu}=G_{\overline \mu \overline \nu}=
G_{\mu \overline \nu}+G_{ \nu \overline \mu}=0,  
$$
and that $G_{\mu \mu}$ $G_{\overline \mu \overline \mu} $, $
G_{\mu \overline \mu}$ do not depend upon $\mu$. Thus these consistency conditions are equivalent
to the symmetrized dynamical equations \ref{symdyn}.  
\section{Hermiticity conditions for superfields:}
We take the gauge group to be $SU(N)$. Then the physical fields $X_m$ and $\phi^\alpha$ are
anti-hermitian matrices. At this point, we need to derive the associated hermiticity conditions
for our superfields $A_m$, $A_\alpha$. Consider, in general a superfield 
\beq
F(\underline x, \underline \theta)= \sum_{p=0}^{16}\sum_{\alpha_1,\ldots,
\alpha_p} {\theta^{ \alpha_1}\cdots
\theta^{ \alpha_p}\over p !}F^{[p]}_{\alpha_1\ldots \alpha_p}(\underline x), 
\label{exp}
\eeq
Then 
$$
F^\dagger (\underline x, \underline \theta)= \sum_{p=0}^{16}\sum_{\alpha_1,\ldots,
\alpha_p}  F^{[p]\dagger }_{\alpha_1\ldots \alpha_p}(\underline x){\theta^{ \alpha_p \dagger}\cdots
\theta^{ \alpha_1 \dagger }\over p !}, 
$$
If $F=F_b$ is bosonic,   $F_b^{[p]}$ is commuting (resp. anticommuting) for $p$ even (resp. $p$
odd). Then, assuming that $\theta^{\alpha \dagger}=\theta^{\alpha } $, we may write 
$$
F_b^\dagger (\underline x, \underline \theta)=K_b \sum_{p=0}^{16}\sum_{\alpha_1,\ldots,
\alpha_p}  {\theta^{ \alpha_1}\cdots
\theta^{ \alpha_p }\over p !}F^{[p]\dagger }_{b \alpha_1\ldots \alpha_p}(\underline x) K_b
$$ 
where
\beq
K_b=\left(-1\right)^{{\cal R}({\cal R}+1)/2}. 
\label{Kbdef}
\eeq 
where 
\beq
{\cal R}=\theta^\alpha \partial_\alpha
\label{Rdef}
\eeq
If $F=F_f$ is fermionic,   $F_f^{[p]}$ is anticommuting (resp. commuting) for $p$ even (resp. $p$
odd). Then, 
$$
F_f^\dagger (\underline x, \underline \theta)= K_f \sum_{p=0}^{16}\sum_{\alpha_1,\ldots,
\alpha_p} {\theta^{ \alpha_1}\cdots
\theta^{ \alpha_p }\over p !}F^{[p]\dagger }_{f \alpha_1\ldots \alpha_p}(\underline x)
K_f, 
$$ \beq 
K_f=\left(-1\right)^{{\cal R}({\cal R}-1)/2}. 
\label{Kfdef}
\eeq
One may verify that the superfields $A_m$, $A_\alpha $ have  decomposition of the type
\ref{exp} with 
$F^{[p]\dagger }_{\alpha_1\ldots \alpha_p}=-F^{[p]}_{\alpha_1\ldots \alpha_p}$ for all $p$. 
Thus we conclude that $A_m^\dagger=-K_b A_m K_b$,  $A_\alpha^\dagger=-K_f A_\alpha K_f$. 
Next consider the effect of the superderivative operator. The action on the $p$th component of a
superfield \ref{exp} is given by
$$
\left(D_{\alpha} F\right)^{[p]}_{\alpha_1,\ldots, \alpha_p}
=F^{[p+1]}_{\alpha\, \alpha_1\ldots \alpha_p}-
\sum_{i=1}^p\left(-1\right)^{i+1}\sigma^m_{\alpha, \alpha_i}
\partial_m F^{[p-1]}_{\alpha_1\ldots /\!\!\!\! \alpha_i \ldots  \overline
\alpha_p}
$$
Since the matrix $\sigma^m$ are real, we immediately get 
\beq
D_{\mu} \left (K_b F_b^\dagger K_b \right)=K_f \left(D_{\mu} F_b\right)^\dagger K_f,\quad 
D_{\mu} \left (K_b F_f^\dagger K_b \right)=K_f \left(D_{\mu} F_f\right)^\dagger K_f
\label{hermvr}
\eeq
The last equations are of course consistent with the fact that the superderivatives transform a
bosonic superfield into a fermionic one and vice versa. At this time, the fact that $A_\alpha$ and
its superderivatives satisfy different hermiticity conditions leads to complications which we will
avoid by only looking at solutions such that $\phi^\alpha=0$. For these purely bosonic solutions
$A^{[2p]}_{\alpha,\, \alpha_1,\ldots \alpha_{2p}}=0$ and $A^{[2p+1]}_{m,\, \alpha_1,\ldots
\alpha_{2p+1}}=0$. All superfield components are commuting, and we may choose, instead of the
above, 
\beq
K_b=K_f=K=\left(-1\right)^{{\cal R}({\cal R}-1)/2}. 
\label{Kdef}
\eeq 
Then, it is easy to show that $\Psi(\lambda)$ and   $\left(K\Psi^\dagger(1/\lambda^*) K\right)^{-1}$ satisfy
the same equation. Thus we shall assume that 
\beq
\Psi(\lambda)=K\Psi^{\dagger -1}(1/\lambda^*) K
\label{hermcond}
\eeq 
\section{The two pole ansatz}
As for self-dual Yang-Mills in four dimensions, we assume that $\Psi$ is a meromorphic function 
of $\lambda$. Condition \ref{hermcond} shows that poles appear in pairs. The simplest ansatz involve
two poles.    The following displays the corresponding solution, for the gauge group $SU(2)$,
following the line of ref\cite{BZ78}  rather closely. Taking the poles at zero and $\infty$ we
write the ansatz  
\begin{eqnarray}
\Psi(\lambda)&=&\left(u {\bf 1}+\lambda fA-
{ \tilde f\tilde A\over \lambda}\right)\nonumber\\
\Psi^{-1}(\lambda)&=&\left(u {\bf 1}-\lambda fA+
{ \tilde f\tilde A\over \lambda}\right)
\label{ansBZ}
\end{eqnarray}
where 
\beq
A={1\over a \tilde a+b \tilde b}\left(\begin{array}{cc}
ab &a^2 \\
-b^2 &-ab 
\end{array}\right). 
\label{Adef}
\eeq
In these definitions $u$, $f$, $a$, $b$ are superfields. In agreement with equations \ref{Kdef},
we introduce the notation 
\beq
\tilde F= KF^\dagger K
\label{tidef}
\eeq
for any (matrix valued or not) superfield.  It is easy to see that 
\beq
A^2=\tilde A^2=0,\quad 
\left[A,\, \tilde A\right]_+={\bf 1}. 
\label{Aprop}
\eeq
The equations just written are such that  the definitions \ref{ansBZ} are
consistent with equation \ref{hermcond}, and with the relation $\Psi(\lambda)
\Psi^{-1}(\lambda)=1$, provided we assume that 
\beq
u^2=1-f\tilde f. 
\label{udef}
\eeq
Next, we derive algebraic equations for the superfields appearing in the ansatz, by rewriting
equation \ref{BZ}  as 
\beq
B_{\mu}+\lambda B_{\overline\mu}=
\Psi(\lambda)\left(\Delta_{\mu}+\lambda \Delta_{\overline
\mu}\right)\Psi^{-1}(\lambda) . 
\label{Beq1}
\eeq
Identifying the powers in $\lambda$ gives the following set of 
independent equations
\begin{eqnarray}
\tilde f \tilde A\Delta_\mu\left(\tilde f \tilde A\right)&=&0 
\label{class1}\\
 \tilde f\tilde A\Delta_{\mu}u
- u \Delta_{\mu} \left(\tilde f\tilde A\right)
- \tilde f\tilde A  \Delta_{\overline \mu}\left(\tilde f\tilde A\right)&=&0
\label{class2}\\
  u {\bf 1}\Delta_{\mu}u
+   \tilde f\tilde A   \Delta_{\overline \mu}u 
+ \tilde f\tilde A\Delta_{\mu} \left(fA\right)
- u \Delta_{\overline \mu}\left(\tilde f\tilde A\right)
+ fA\Delta_{\mu} \left(\tilde f\tilde A\right) &=&-B_\mu,  
\label{class3} 
\end{eqnarray}
together with three more relations deduced from the above according to equation \ref{hermcond}. 
At this point it is useful to write 
\beq
A={1\over a \tilde a+b\tilde b} \Upsilon. 
\label{Upsdef}
\eeq
Since the matrix $\Upsilon$ is such that $\Upsilon^2=0$. Equation \ref{class1} is satisfied iff 
\beq
\Delta_\mu\tilde a=\Delta_\mu \tilde b=0.  
\label{eq1}
\eeq
Equation \ref{class2} may be transformed into
$$
\tilde \Upsilon \Delta_{\mu}\tilde g 
=  \tilde
\Upsilon  \Delta_{\overline \mu}
 \tilde \Upsilon 
$$
where we have let 
\beq
\tilde g=u {a \tilde a+b\tilde b \over \tilde f } 
\label{gdef}
\eeq
Equation \ref{class2} is satisfied if we have 
\beq
\Delta_\mu \tilde g=\tilde h_{\overline \mu }, \quad 
\tilde h_{\overline \mu }=\tilde b \Delta_{\overline
\mu }\tilde a -\tilde a\Delta_{\overline \mu }\tilde b. 
\label{eq2}
\eeq
Remarkably, equation \ref{eq1} is a particular case of equations which already appeared in
ref\cite{GS98} where  general solutions were obtained which are only dependent upon $x^0$ and
$x^9$. We shall obtain solutions of equations \ref{eq2} below. Once these two equations are solved,
equation
\ref{class3} allows to derive the vector potentials. For this it is convenient to rewrite it under
the form 
\beq
B_\mu=            {1\over u}\Delta_{\mu}u
+            {\tilde \Upsilon \over \tilde g}\Delta_{\mu} \left({\Upsilon \over  g}\right)
-              \Delta_{\overline \mu}\left({\tilde \Upsilon \over \tilde g}\right)
+             {\Upsilon \tilde \Upsilon\over  g}\Delta_{\mu} \left({1 \over \tilde g}\right) 
\label{Beq}
\eeq
\section{A particular solution}
At this preliminary stage, and in order to arrive at a concrete solution, we choose a simple
particular ansatz. We only retain dependence in $x^0\equiv t$ and $x^9\equiv x$. A simple linear
solution of equations \ref{eq1} is 
\beq
a=1, \quad b=t+i\sum_\mu\theta^\mu \theta^{\overline \mu}, 
\label{s1}
\eeq
so that 
$$
\Delta_\mu a=0, \quad \Delta_\mu b=2 D_\mu b=\left(\theta^\mu+i\theta^{\overline \mu}\right) 
$$
$$
a\tilde a+b\tilde b=b  +\tilde b  =2t. 
$$
Then equation \ref{eq2} gives 
\beq
g=-8 x +c
\label{s2}
\eeq
where $\Delta_{\overline \mu}c=0$. We will simply choose $c$ to be a constant. Using equations
\ref{udef}, \ref{gdef},  we obtain 
$$
u =\sqrt{ {\left|c-8x\right|^2  \over 4t^2+\left|c-8x\right|^2} }. 
$$
Finally, using equation \ref{Beq} one gets 
$$
B_\mu=(\theta^\mu+i\theta^{\overline \mu}) \left \{ 
  {4t \over \left(4t^2+\left|c-8x\right|^2\right)} 
-  {2 \over \left| (c-8x)\right|^2} \left(\begin{array}{cc}
 \tilde b  +2 \tilde b  ^2  b & \tilde b  ^2  \\
  1  +2 \tilde b b &  \tilde b
\end{array}\right)\right. 
$$
$$
 \left. + 
 {8\over (c^*-8x)^2  } 
\left(\begin{array}{cc}
b&1 \\
-b  ^2 &-b
\end{array}\right) \right \} 
+(\theta^\mu-i\theta^{\overline \mu})  \left \{ 
{-16 \left(16 x-c-\tilde c\right)t^2 \over \left|c-8x\right|^2 \left(4t^2+\left|c-8x\right|^2
\right)}\right.
$$
$$
\left.  
-  {8 \over (c-8x)\left| (c-8x)\right|^2 } 
\left(\begin{array}{cc}
 \tilde b b  +\tilde b  ^2  b  ^2 &\tilde b+\tilde b  ^2  b\\
  b+\tilde b b  ^2 & 1 +  \tilde bb  
\end{array}\right)
+ {2\over c^*-8x}
\left(\begin{array}{cc}
1 &0 \\
-2\tilde  b&-1 
\end{array}\right)
\right.
$$
\beq 
\left.
 - 
{8\over (c^*-8x)\left| (c-8x)\right|^2} 
\left(\begin{array}{cc}
b  \tilde b+ 1 &-b\tilde b  ^2-   \tilde b\\
- b  ^2 \tilde b-b  & b  ^2  \tilde b  ^2 + b \tilde b
\end{array}\right)\right\}
\label{simple}
\eeq

\section{Outlook}
It seems clear that the symmetrised system of equations \ref{symdyn} is completely and explicitly
integrable much like self-dual Yang-Mills in four dimensions. Note that, in the gauge introduced in
ref.\cite{GS98} where $A_{\overline \mu}=0$, the right most equations  \ref{symdyn} give
$D_{\overline \mu}A_{\nu}+D_{\overline \nu}A_{\mu}=0 $, for $\mu\not=\nu$. This is precisely the
condition which was used in ref\cite{GS98} to let $A_{\mu}=D_{\overline \mu}\Phi$. In other words,
the present Lax pair is  equivalent to the set of equations which was  previously solved in
ref.\cite{GS98}. 

Concerning the full Yang-Mills equations or equivalently the unsymmetrised equations
\ref{dynuu}--\ref{dynud}, any solution is also a solution of the symmetrised equations
\ref{symdyn}. Thus we should be able to derive solutions of the latter which are general enough so
that we may impose that they be solutions of the former. This problem is currently under
investigation. 
\bigskip 

\noindent { \bfseries\Large Acknowledgements: }\\
It is a pleasure to acknowledge stimulating discussions with P. Forgacs and H.~Samtleben.

\nom\modification\fin